\documentclass[twocolumn,preprintnumbers,amsmath,aps]{revtex4}
\usepackage{graphicx}
\usepackage{dcolumn}
\usepackage{bm}
\begin{document}
\newcommand{\kvec}{\mbox{{\scriptsize {\bf k}}}}
\def\eq#1(\ref{#1)}
\def\fig#1{\ref{#1}}
\title{Superconducting properties of under- and over-doped Ba$_{x}$K$_{1-x}$BiO$_{3}$ perovskite oxide}
\author{D. Szcz{\c{e}}{\'s}niak$^{1}$, A. Z. Kaczmarek$^{2}$, R. Szcz{\c{e}}{\'s}niak$^{1, 2}$, S. V. Turchuk$^{1, 3}$, H. Zhao$^{1, 4}$, E. A. Drzazga$^{2}$}
\affiliation{1. Institute of Physics, Jan D{\l}ugosz University in Cz{\c{e}}stochowa, Ave. Armii 
Krajowej 13/15, 42200 Cz{\c{e}}stochowa, Poland}\email{d.szczesniak@ajd.czest.pl}
\affiliation{2. Institute of Physics, Cz{\c{e}}stochowa University of Technology, Ave. Armii Krajowej 19, 42200 Cz{\c{e}}stochowa, Poland}
\affiliation{3. Department of Physics, Lesya Ukrainka East European National University,\\ Ave. Volya 13, 43000 Lutsk, Ukraine}
\affiliation{4. Institute for Molecules and Materials UMR 6283, Le Mans University, Ave. Olivier Messiaen, 72085 Le Mans, France.}
\date{\today} 
\begin{abstract}
In the present study, we investigate the thermodynamic properties of the Ba$_{x}$K$_{1-x}$BiO$_{3}$ (BKBO) superconductor in the under- ($x=0.5$) and over-doped ($x=0.7$) regime, within the framework of the Migdal-Eliashberg formalism. The analysis is conducted to verify that the electron-phonon pairing mechanism is responsible for the induction of the superconducting phase in the mentioned compound. In particular, we show that BKBO is characterized by the relatively high critical value of the Coulomb pseudopotential, which changes with doping level and does not follow the Morel-Anderson model. In what follows, the corresponding superconducting band gap size and related dimensionless ratio are estimated to increase with the doping, in agreement with the experimental predictions. Moreover the effective mass of electrons is found to take on high values in the entire doping and temperature region. Finally, the characteristic dimensionless ratios for the superconducting band gap, the critical magnetic field and the specific heat for the superconducting state are predicted to exceed the limits set within the Bardeen-Cooper-Schrieffer theory, suggesting pivotal role of the strong-coupling and retardation effects in the analyzed compound. Presented results supplement our previous investigations and account for the strong-coupling phonon-mediated character of the superconducting phase in BKBO at any doping level.
\end{abstract}
\maketitle

\section{INTRODUCTION}

The Ba$_{x}$K$_{1-x}$BiO$_{3}$ (BKBO) perovskite oxide constitutes one of the most extensively analyzed superconductors, up to date \cite{batlogg, cava, uemura, hinks, graebner, navarro1, pei, shirai, loong, samuely, baumert, braden, navarro2, barilo, nazia, zhao1, khosroabadi, liu}. The reason for such a considerable interest in this compound stems from the relatively high critical temperature ($T_{c}$) values which can be obtained in this compound \cite{batlogg, pei, lee}, as well as the non-conventional behavior of its other thermodynamic properties \cite{navarro1, navarro2, nazia, szczesniak1}. In this context, considerable attention was given to the understanding of pairing mechanism in the discussed material, which appeared to be a formidable challenge for both theory and experiment. By comparison to the sibling BaPb$_{0.75}$Bi$_{0.25}$0$_{3}$ (BPBO) \cite{sleight}, initially it was suggested that BKBO exhibits non-phononic pairing mechanism \cite{batlogg}. Specifically, this observation was made within the analysis of the physical properties such as the density of states and the isotope effect \cite{batlogg}. However, later investigations predicted that, although BKBO does not behave like most of the phonon-mediated superconductors, its superconducting phase is governed by the high-frequency modes and discussed material should be considered as a conventional superconductor \cite{braden}. This conclusion was additionally reinforced by the lack of the cooperative magnetic behavior in the discussed perovskite oxide \cite{uemura}.

In favor of the latter predictions, recently we have accounted for the phonon-mediated character of the superconducting phase induced in the optimally doped BKBO {\it i.e} when $x=0.6$ \cite{szczesniak1}. The recalled analysis was conducted within the Eliashberg formalism \cite{eliashberg} - a generalization of the Bardeen-Cooper-Schrieffer (BCS) theory \cite{bardeen1, bardeen2} - to predict the most important thermodynamic properties that describe the superconducting state. Obtained estimates were found to be in good agreement with available experimental results and additionally suggested that previously observed deviations from the BCS scenario may be explained by the strong coupling of electrons and phonons. However, it is important to note that those predictions are valid only for one particular doping case, whereas superconducting phase in BKBO spans doping region from $x=0.5$ to $x=0.7$ \cite{pei, lee, nazia}. Naturally, one may expect changes in the values of the thermodynamic parameters when going from optimal doping towards under- ($x=0.5$) or over-doped ($x=0.7$) regions. Therefore, it is vital to conduct similar test of pari§ng mechanism for the mentioned doping extrema, to supplement our previous results and prove that the general character of the superconducting state in BKBO is preserved despite doping level. Such calculations are also important for the complementary verification of the electron-phonon coupling strength across the entire doping range of the superconducting phase in BKBO.

In this context, herein we use the same methodology as in \cite{szczesniak1} and investigate behavior of the superconducting phase in BKBO at the boundaries of its existence. In particular, two mentioned doping cases ($x=0.7$ and $x=0.5$) are discussed in the framework of the isotropic Eliashberg equations, assuming that critical temperature values are equal to the experimental data presented in \cite{pei, lee}. Herein, special attention is given to the analysis of the electron depairing correlations, the order parameter of the superconducting phase, and the effective mass of electrons. These characteristics are supplemented by the discussion of the specific heat and the thermodynamic critical field. To verify our assumptions on the pairing mechanism in BKBO, obtained results are be compared to the previously determined values of the thermodynamic parameters for $x=0.6$, but also to the other available data.

\section{NUMERICAL RESULTS}

In the present study, the thermodynamic properties of the BKBO superconductor are numerically analyzed within the formalism of the Eliashberg equations \cite{eliashberg}. The choice of the main theoretical technique is motivated by the relatively high value of the the electron-phonon coupling constant ($\lambda$), which is much greater than 0.5 \cite{carbotte, cyrot} for both Ba-doping cases; explicitly $\lambda=1.1$ for $x=0.5$ and $\lambda=1.31$ for $x=0.7$ \cite{nazia}. In particular, the Eliashberg equations are solved here on the imaginary axis and later analytically continued on the real axis to obtain quantitative estimates of the selected properties. Herein, the mentioned equations are treated within the numerical procedures presented in \cite{durajski1, durajski2, jarosik1, jarosik2}. The exact form of the Eliashberg equations on the imaginary axis for our numerical computations reads \cite{szczesniak2}:
\begin{equation}
\label{eq1}
\phi_{n}=\frac{\pi}{\beta}\sum_{m=-M}^{M}
\frac{K\left(i\omega_{n}-i\omega_{m}\right)-\mu^{\star}\theta\left(\omega_{c}-|\omega_{m}|\right)}
{\sqrt{\omega_m^2Z^{2}_{m}+\phi^{2}_{m}}}\phi_{m},
\end{equation}
\begin{equation}
\label{eq2}
Z_{n}=1+\frac{1}{\omega_{n}}\frac{\pi}{\beta}\sum_{m=-M}^{M}
\frac{K\left(i\omega_{n}-i\omega_{m}\right)}{\sqrt{\omega_m^2Z^{2}_{m}+\phi^{2}_{m}}}
\omega_{m}Z_{m},
\end{equation}
where the first of the above equations describe the order parameter function ($\phi_{n}\equiv\phi\left(i\omega_{n}\right)$), whereas the second one describes the wave function renormalization factor ($Z_{n}\equiv Z\left(i\omega_{n}\right)$). Moreover, in Eqs. (\ref{eq1}) and (\ref{eq2}), $i$ denotes the imaginary unit and $\omega_{n}$ is the $n$-th Matsubara frequency where $\omega_{n}\equiv \left(\pi/\beta\right)\left(2n-1\right)$ and $\beta\equiv\left(k_{B}T\right)^{-1}$, whereas $k_{B}$ is the Boltzmann constant. In this context the $\omega_{c}$ is the cut-off frequrency, which we assume to be equal to $10\Omega_{\rm max}$, where $\Omega_{\rm max}$ is the maximum phonon frequency and equals to $62.99$ meV and $63.99$ meV for $x=0.5$ and $x=0.7$, respectively. Further, $\mu^{\star}$ models the electron depairing correlations and is known as the Coulomb pseudopotential, $\theta$ is simply the Heaviside function, and $K\left(z\right)$ describes the pairing kernel given as:
\begin{equation}
\label{eq3}
K\left(z\right)\equiv 2\int_0^{+\infty}d\omega\frac{\omega}{\omega ^2-z^{2}}\alpha^{2}\left(\omega\right)F\left(\omega\right).
\end{equation}
In Eq. (\ref{eq3}) $\alpha^2(\omega)F(\omega)$ is the Eliashberg spectral functions, where $\alpha^2(\omega)$ denotes effective electron-phonon coupling function, $F(\omega)$ stands for the phonon density of states and $\omega$ describes the phonon frequency. The $\alpha^2(\omega)F(\omega)$ is essential in our calculations, since it models the electron-phonon interactions and enters the Eliashberg equation as a main input parameter. In this study, we use two forms of the Eliashberg function, one for each of the considered Ba-doping cases. Both of the functions has been calculated in \cite{nazia} from the first-principles. To this end, the numerical stability is obtained by taking into account 2201 Matsubara frequencies, for $T \geq T_{0}$, where $T_{0}=2$ K.

The aforementioned Eliashberg equations on the imaginary axis give the estimates of the following thermodynamic properties:  the order parameter of the superconducting state ($\Delta_{n}=\phi_{n}/Z_{n}$), the critical value of the Coulomb pseudopotential ($\mu_{c}$), the free energy difference between the normal and superconducting state ($\Delta F$), the thermodynamic critical field ($H_{c}$), and the specific heat difference between the normal and superconducting state ($\Delta C$).

As already mentioned, the imaginary axis Eliashberg equations are next analytically continued on the real axis ($\Delta_{n}\rightarrow\Delta\left(\omega\right)$). Specifically this is done for the order parameter function by using the Pad{\' e} analytical continuation method \cite{beach,marsiglio}, given as:
\begin{equation}
\label{eq4}
\Delta\left(\omega\right)=\frac{p_{\Delta 1}+p_{\Delta 2}\omega+...+p_{\Delta r}\omega^{r-1}}
{q_{\Delta 1}+q_{\Delta 2}\omega+...+q_{\Delta r}\omega^{r-1}+\omega^{r}},
\end{equation}
where $p_{\Delta j}$ and $q_{\Delta j}$ denote coefficients which take integer values and $r=550$. In what follows the order parameter function on the real axis can be written as \cite{eliashberg}, \cite{carbotte}:
\begin{equation}
\label{eq5}
\Delta\left(T\right)={\rm Re}\left[\Delta\left(\omega=\Delta\left(T\right),T\right)\right].
\end{equation}
Above equation give quantitative estimates of the superconducting energy band gap at the Fermi level ($\Delta_{g}=2\Delta\left(0\right)$ where $\Delta\left(0\right)\simeq\Delta\left(T_{0}\right)$) and the related properties of interest.

\begin{figure}[ht]
\centering
\includegraphics[width=\columnwidth]{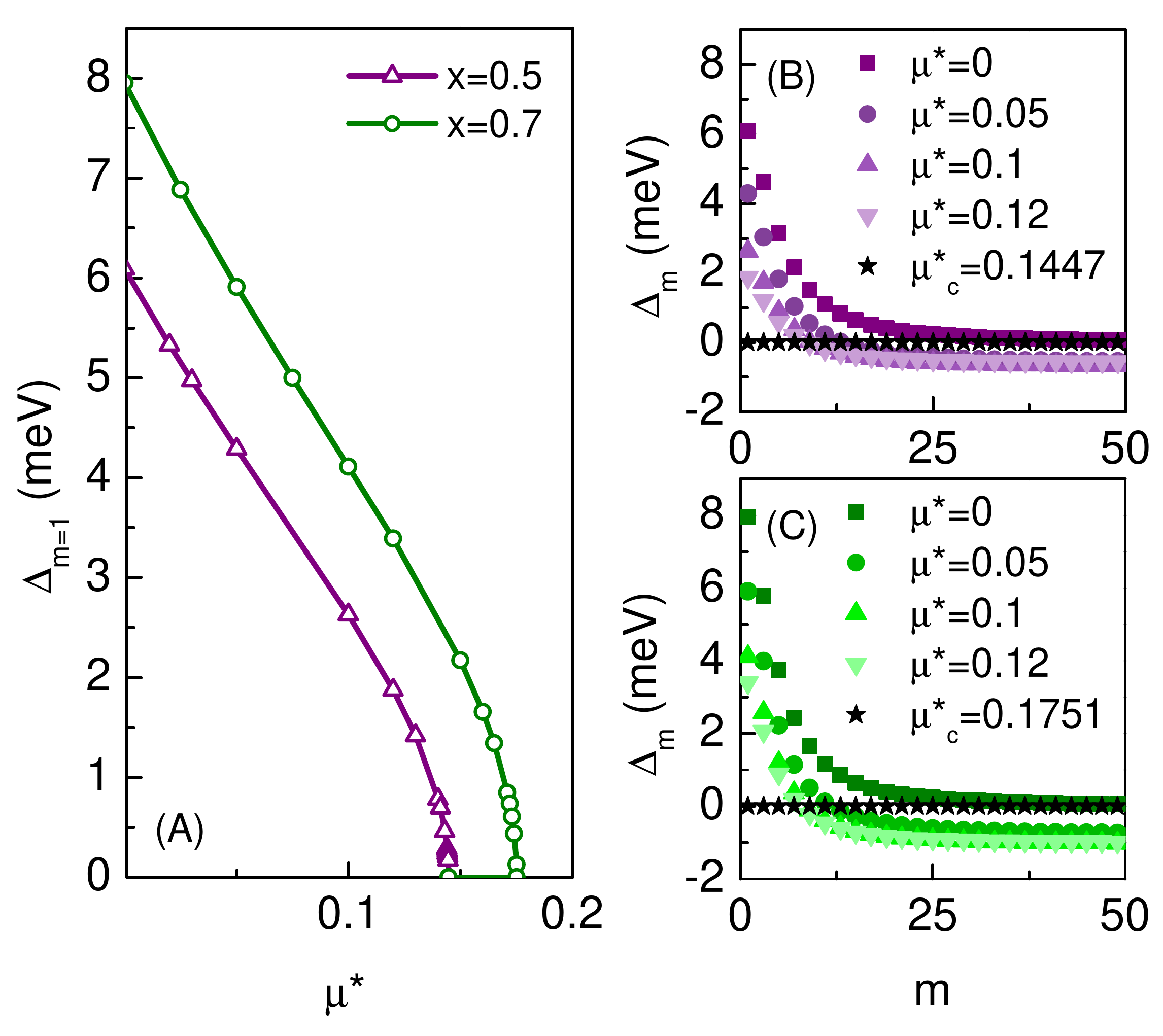}
\caption{(A) The maximum value of order parameter ($\Delta_{m=1}$) as a function of Coulomb pseudopotential ($\mu^{*}$) for under- and over-doped BKBO. The order parameter ($\Delta_{m}$) as a function of $m$ for under- (B) and over-doped (C) BKBO; only selected values of $\mu^{*}$ are presented for better clarity.}
\label{fig01}
\end{figure}

The Eliashberg equations on the imaginary axis, given by Eqs. (\ref{eq1}) and (\ref{eq2}), yield set of the temperature-dependent $\phi_{n}$ and $Z_{n}$ functions. As already mentioned these functions can be used to obtain full thermal characteristic of the order parameter on the imaginary axis ($\Delta_{n}=\phi_{n}/Z_{n}$). In this context, it is instructive to first analyze the value of the Coulomb pseudopotential in the terms of the $\Delta_{n}$ function. Note that $\mu^{\star}$ will have critical value ($\mu^{\star}_{c}$) at the point where $\Delta_{n}$ function is equal to zero. This corresponds to the metal-superconductor phase transition point and marks the physically-relevant value of the Coulomb pseudopotential. Herein, the $\mu^{\star}_{c}$ is determined by solving Eqs. (\ref{eq1}) and (\ref{eq2}) for the different values of $\mu^{\star}$, with the assumption that $T=T_{c}$, where $T_{c}$ is equal to 22.5 K and 28.3 K for $x=0.5$ and $x=0.7$, respectively. We remind that assumed values of $T_{c}$ are taken from experimental predictions given in \cite{pei, lee}, as their averages.

In Fig. \ref{fig01} (A), we present obtained functional behavior of the maximum value of the order parameter ($\Delta_{m=1}$ {\it i.e.} the value for the first Matsubara frequency $m=1$) on $\mu^{\star}$ for both cases of Ba-doping. Therein, Figs.  \ref{fig01} (B) and (C) present dependence of the order parameter function on $m$ and show saturation of the solutions for $m>25$, what assures that the numerical accuracy of the conducted calculations is sufficiently high.

These results allow to determine the desired $\mu^{\star}_{c}$ value, which is equal to 0.14 and 0.18 for the $x=0.5$ and $x=0.7$ case, respectively. We observe, that the predicted $\mu^{\star}_{c}$ values are relatively high comparing to the typical phonon-mediated superconductors, however they are very close to the previously predicted estimates for the optimally doped BKBO with $x=0.6$ \cite{szczesniak1}. Similarly to \cite{szczesniak1}, the observed high values of the Coulomb pseudopotential does not follow the Morel-Anderson model \cite{morel}, what can be explained by the small influence of the retardation effects on its value, as suggested within the approach of Bauer {\it et al.} \cite{bauer}. Nonetheless, herein we prove that BKBO is likely to present high values of the $\mu^{\star}_{c}$ parameter in the entire range of the Ba-doping. Therefore, the electron depairing correlations are found to be strong for both considered dopings. Note also, that the presented analysis suggest also decrease of the $\mu^{\star}_{c}$ value with the $x$, in strong contradiction to the usually adopted constant value of $\mu^{\star}$ in calculations based on the approximate analytical models \cite{navarro1, navarro2, nazia}.

\begin{figure}[ht]
\centering
\includegraphics[width=\columnwidth]{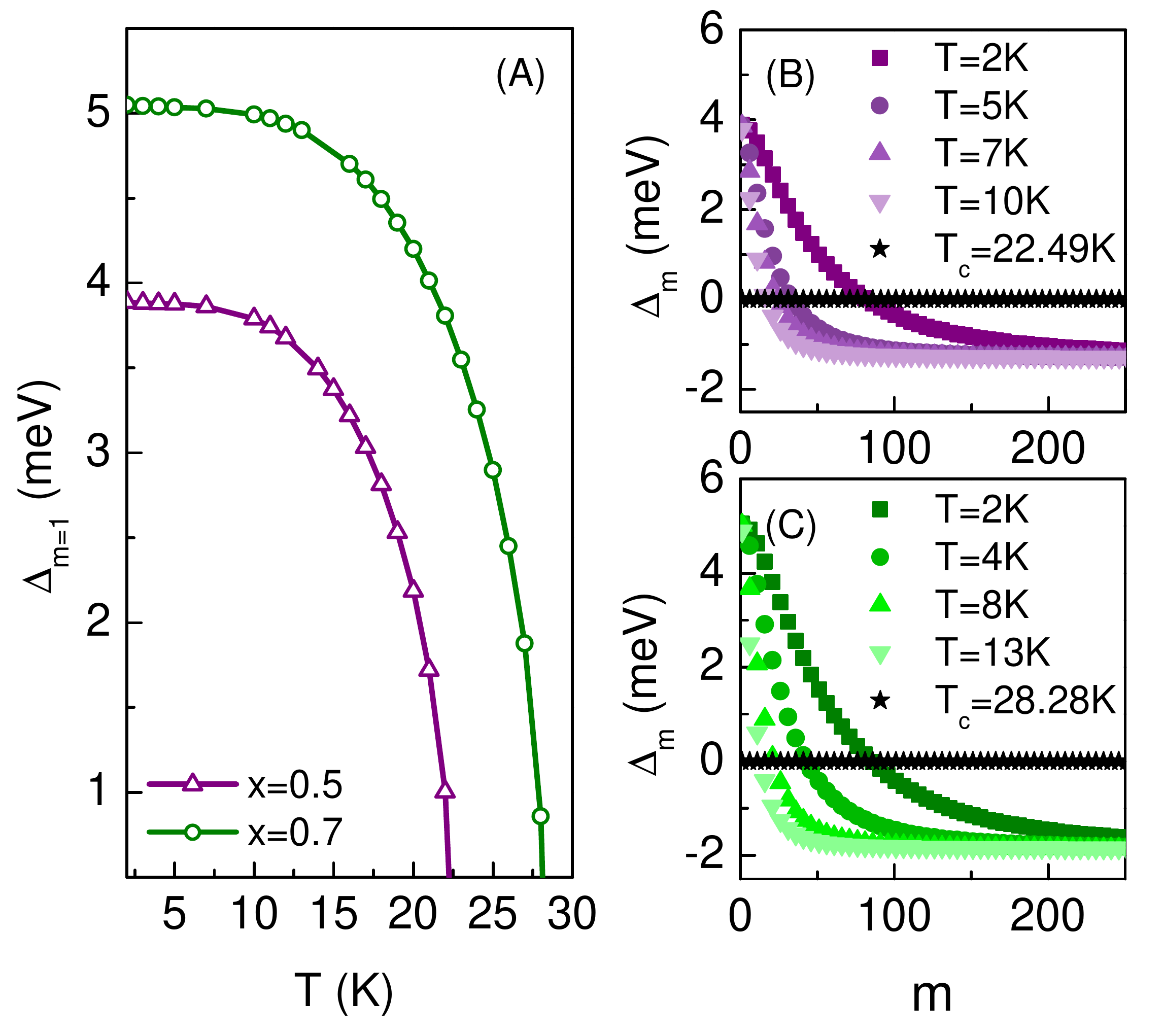}
\caption{(A) The maximum value of the order parameter ($\Delta_{m=1}$) as a function of the temperature for under- and over-doped BKBO. The order parameter ($\Delta_{m}$) as a function of $m$ for under- (B) and over-doped (C) BKBO; only selected values of temperature are presented for better clarity.}
\label{fig02}
\end{figure}

The knowledge of the $\mu^{\star}_{c}$ parameter value allows next to calculate the temperature dependence of the order parameter. In Fig. \ref{fig02} (A), we depict the maximum value of the order parameter ($\Delta_{m=1}$) on the temperature in the range from $T_{0}$ to $T_{c}$ for both considered doping levels. Presented $\Delta_{m=1}(T)$ function exhibits common thermal behavior for the phonon-mediated superconductors \cite{carbotte} {\it i.e.} after plateau at low temperatures, order parameter noticeably decreases together with the increase of the temperature. Moreover, we observe that the higher doping levels of Ba lead to the higher values of the order parameter, what follows increase of the $T_{c}$ along with the doping. Finally, the fact that $\Delta_{m=1}$ takes the value of zero for the assumed experimental value of $T_{c}$, partially proves the correctness of our numerical calculations. This fact is additionally reinforced by the behavior of the order parameter on $m$, as plotted in Figs. \ref{fig02} (B) and (C). Therein, it can be easily noticed that for $m>150$ Eliashberg solutions are suppressed despite the temperature value, meaning that the limit $m=1100$, assumed during numerical calculations, is well above the saturation point.

We note, that the maximum value of the order parameter at $T=T_{0}$ allows to estimate the paramount value of the superconducting band gap $\Delta_{g}=2\Delta_{m=1}(0)$. In the first approximation the $\Delta_{m=1}(0)$ parameter can be taken directly from the imaginary axis solutions and gives $\Delta_{g}$ to be equal to 7.77 meV for $x$=0.5 and 10.10 meV for $x$=0.7. However, the quantitative estimates of the discussed parameter are possible only when analyzed frequencies correspond directly to the quasi-particle energies. To establish such relation the imaginary axis solutions has to be analytically continued on the real axis, as described by Eqs. (\ref{eq4}) and (\ref{eq5}). The corrected values of $\Delta_{g}$ are: 7.90 meV and 10.31 meV for $x$=0.5 and $x$=0.7, respectively. Therefore, the imaginary and real axis predictions are not far from each other and the former approximation proves its predictive capabilities.

To compare our predictions with other estimates available in the literature it is convenient to calculate one of the characteristic ratios, namely $R_{\Delta}\equiv 2\Delta\left(0\right)/k_{B}T_{c}$, where $\Delta\left(0\right)$ denotes the order parameter value at $T_{0}$ obtained from the real axis solutions of the Eliashberg equations. The calculated values of $R_{\Delta}$ ratio are equal to 4.07 and 4.23 for $x=0.5$ and $x=0.7$, respectively. Therefore, value of the $R_{\Delta}$ parameter presents increase together with $x$, in agreement with the experimental predictions in \cite{zhao2}. Moreover, the obtained values itself are very close to the ones estimated from the experiment. In this context, of particular attention are results presented in \cite{zhao2, barilo, samuely}, which presents the most complementary and recent experimental investigations to our knowledge.

\begin{figure}[ht]
\centering
\includegraphics[width=\columnwidth]{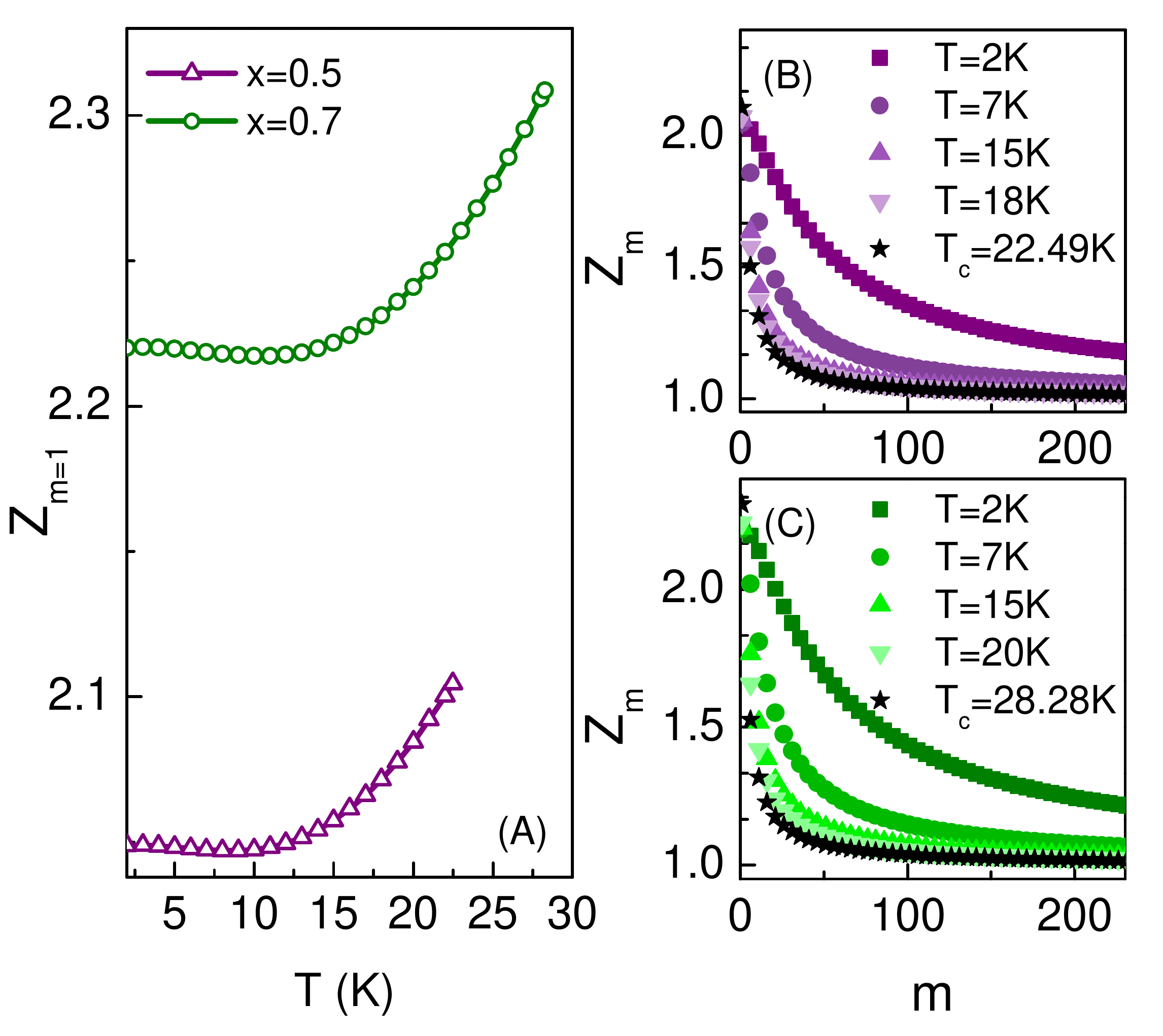}
\caption{(A) The maximum value of wave function renormalization factor ($Z_{m=1}$) as a function of temperature for under- and over-doped BKBO. The wave function renormalization factor ($Z_{m}$) as a function of $m$ for under- (B) and over-doped (C) BKBO; only selected values of temperature are presented for better clarity.}
\label{fig03}
\end{figure}

It is important to note that the imaginary axis solutions of the Eliashberg equations provide other supplementary thermodynamic properties of interest. First of all, the wave function renormalization factor allows to describe with a high accuracy the effective mass of electrons ($m^{\star}_{e}$); $m^{\star}_{e}\simeq Z_{m=1} m_{e}$, where $m_{e}$ denotes the band electron mass and $Z_{m=1}$ is the maximum value of the wave function renormalization factor. In Fig. \ref{fig03} (A) we present the $Z_{m=1}$ parameter as a function of the temperature. It can be easily observed that the $Z_{m=1}$ function increases together with the increase of the temperature and reaches its maximum value at $T=T_{c}$. In this context, we obtain physically-relevant increase of the $m^{\star}_{e}$ with the temperature and estimate the maximum value of the effective mass of electrons to be: $2.10m_{e}$ and $2.31m_{e}$ for $x=0.5$ and $x=0.7$, respectively. Both values are relatively high as for the phonon-mediated superconductors and in agreement with fundamental relation $Z_{m=1}\left(T_{C}\right)=1 + \lambda$ \cite{carbotte}.

\begin{figure}[ht]
\centering
\includegraphics[width=\columnwidth]{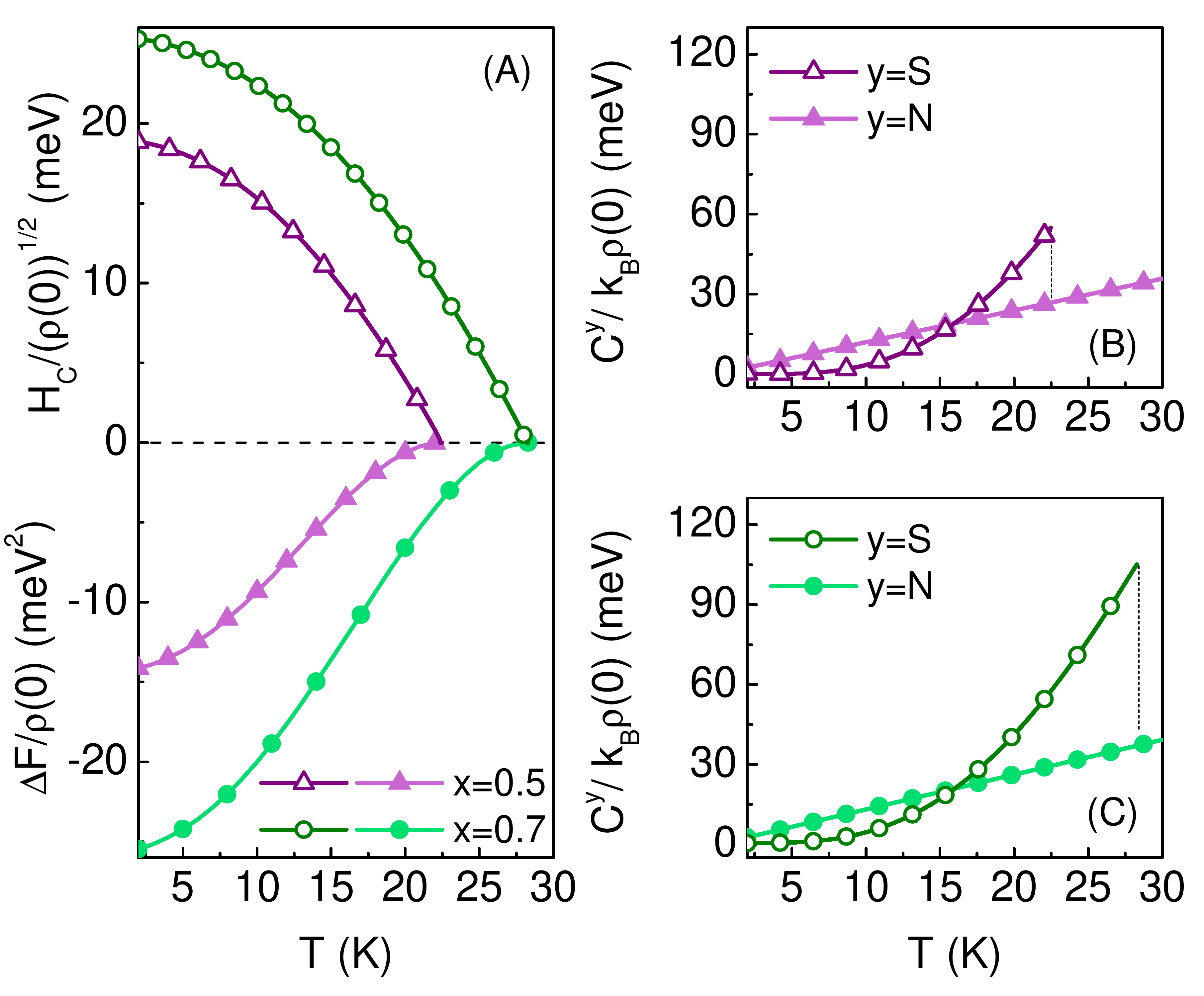}
\caption{(A) The normalized free energy difference between superconducting and normal state ($\Delta F/\rho\left(0\right)$) and normalized critical field ($H_{C}/\sqrt{\rho\left(0\right)}$) as a function of temperature for under- and over-doped BKBO. The thermal behavior of the normalized specific heat for superconducting ($C^{S}$) and normal state ($C^{N}$) as a function of temperature for under- (B) and over-doped (C) BKBO. }
\label{fig04}
\end{figure}

To complement above results it is instructive to investigate two other physical observables such as the thermodynamic critical field ($H_{C}$) and the specific heat of the superconducting state ($C^{S}$). For this reason we first calculate the normalized free energy difference between the superconducting and normal state ($\Delta F/\rho\left(0\right)$) as \cite{bardeen3}:
\begin{eqnarray}
\label{eq6}
\frac{\Delta F}{\rho\left(0\right)}&=&-\frac{2\pi}{\beta}\sum_{m=1}^{M}
\left(\sqrt{\omega^{2}_{m}+\Delta^{2}_{m}}- \left|\omega_{m}\right|\right)\\ \nonumber
&\times&(Z^{{\rm S}}_{m}-Z^{\rm N}_{m}\frac{\left|\omega_{m}\right|}{\sqrt{\omega^{2}_{m}+\Delta^{2}_{m}}}),  
\end{eqnarray}  
In Eq. (\ref{eq6}), $\Delta F$ is the free energy difference between the superconducting (S) and normal (N) state, $\rho\left(0\right)$ denotes the electronic density of states at the Fermi energy, whereas $Z^{\rm S}_{m}$ and $Z^{\rm N}_{m}$ are the the wave function renormalization factors for the superconducting and normal state, respectively. The functional behavior of $\Delta F/\rho\left(0\right)$ parameter is presented in the lower panel of Fig. \ref{fig04} (A) and exhibit negative values, what confirms the stability of the discussed superconducting state.

Next, the $\Delta F/\rho\left(0\right)$ parameter is employed to calculate the normalized thermodynamic critical field ($H_{C}/\sqrt{\rho\left(0\right)}$) according to the below relation \cite{blezius, carbotte}:
\begin{equation}  
\label{eq7}
\frac{H_{C}}{\sqrt{\rho\left(0\right)}}=\sqrt{-8\pi
\left[\Delta F/\rho\left(0\right)\right]}.
\end{equation}
Moreover, with a help of Eq. (\ref{eq6}), the specific heat for the superconducting state ($C^{\rm S}$) is computed from the difference of the specific heat between the superconducting (S) and normal (N) state ($\Delta C$):
\begin{equation}
\label{eq8}
\frac{\Delta C}{k_{B}\rho\left(0\right)}
=-\frac{1}{\beta}\frac{d^{2}\left[\Delta F/\rho\left(0\right)\right]}
{d\left(k_{B}T\right)^{2}}.
\end{equation}
In Eq. (\ref{eq8}), the $\Delta C\equiv C^{\rm S}-C^{\rm N}$ and the specific heat for the normal state ($C^{\rm N}$) reads:
\begin{equation}
\label{eq9}
\frac{C^{N}}{ k_{B}\rho\left(0\right)}=\frac{\gamma}{\beta}.
\end{equation}
where $\gamma$ is the Sommerfeld constant.

The determined $H_{C}/\sqrt{\rho\left(0\right)}$ dependence on temperature is depicted in the upper panel of Fig. \ref{fig04} (A). On the other hand, the specific heat for the superconducting and normal state is presented in Figs. \ref{fig04} (B) and (C) for $x=0.5$ and $x=0.7$, respectively. We note, that all functions presented in Figs. \ref{fig04} (A), (B) and (C) exhibit behavior expected for the phonon-mediated superconductors \cite{carbotte}. An especially characteristic effect can be observed for the $C^{\rm S}$ functions which present distinctive {\it jump} at $T=T_{c}$. Nonetheless, results given in Figs. \ref{fig04} make it possible to calculate two remaining dimensionless ratios, present in the BCS theory. Specifically, the aformentioned ratios can be written as: $R_{\rm H}\equiv T_{c}C^{N}\left(T_{c}\right)/H^{2}_{C}\left(0\right)$ and $R_{\rm C}\equiv \Delta C\left(T_{c}\right)/C^{N}$. For the discussed superconductor the former ratio ($R_{\rm H}$) equals to: $0.146$ and $0.141$ for $x=0.5$ and $x=0.7$, respectively. To this end, the latter ratio ($R_{\rm C}$) presents following values: $2.13$ for $x=0.5$ and $2.31$ for $x=0.7$. Unfortunately, to our knowledge, no experimental predictions on $R_{\rm H}$ and $R_{\rm C}$ has been reported yet. In this context, calculated above values should serve as a reference for corresponding future experimental investigations. 

\section{SUMMARY AND CONCLUSIONS}

In summary, we have provided complementary analysis of the most important thermodynamic parameters of the superconducting state in the under- ($x=0.5$) and over-doped ($x=0.7$) Ba$_{x}$K$_{1-x}$BiO$_{3}$ (BKBO) compound. The analysis was conducted within the Eliashberg formalism \cite{eliashberg} - a strong-coupling generalization of the BCS theory \cite{bardeen1, bardeen2} - to account for the phonon-mediated character of the superconducting phase in the discussed material.

Specifically, our analysis showed that BKBO is characterized by the following critical values of the Coulomb pseudopotential ($\mu^{*}_{c}$) {\it i.e.} 0.14 and 0.18 for $x=0.5$ and $x=0.7$, respectively. We have noted, that both calculated $\mu^{*}_{c}$ values are relatively high as for the phonon-mediated superconductors \cite{carbotte} and does not directly follow the Morel-Anderson model \cite{morel}, similarly as it was previously found for the BKBO with $x=0.6$ \cite{szczesniak1}. On the other hand, observed high values of $\mu^{*}_{c}$ parameter may be explained when appropriate corrections of Bauer {\it et al.} \cite{bauer} are applied to the Morel-Anderson model. Within this explanation retardation effects are predicted to have small influence on the Coulomb pseudopotential, while superconducting phase is still mediated by the phonons. To further verify coupling mechanism in the discussed compound, determined $\mu^{*}_{c}$ parameters where used to calculate other characteristic physical observables.

In what follows we have analyzed the thermodynamic properties such as the superconducting energy band gap, the critical magnetic field and the specific heat for the superconducting state, which next served for the estimation of the corresponding characteristic dimensionless ratios, familiar in the BCS theory. First of the mentioned ratios, defined as $R_{\Delta}\equiv 2\Delta\left(0\right)/k_{B}T_{c}$, was estimated to be equal to 4.07 and 4.23 for $x=0.5$ and $x=0.7$, respectively. We note that this value greatly exceeds predictions of the BCS theory, which suggest $R_{\Delta}=3.53$. Moreover, the increase of $R_{\Delta}$ parameter (and the related superconducting energy band gap) with the doping, was found to be in agreement with the corresponding experimental predictions \cite{zhao2}. Similarly, two remaining parameters, namely $R_{\rm H}\equiv T_{c}C^{N}\left(T_{c}\right)/H^{2}_{C}\left(0\right)$ and $R_{\rm C}\equiv \Delta C\left(T_{c}\right)/C^{N}$, present values which notably differ from the estimates of the BCS theory. In particular, $R_{\rm H}$ equals to $0.146$ for $x=0.5$ and $0.141$ for $x=0.7$, whereas $R_{\rm C}$ equals to $2.13$ and $2.31$ for $x=0.5$ and $x=0.7$, respectively. Note that for $R_{\rm H}$ and $R_{\rm C}$, the BCS theory gives values of 0.168 and 1.43, respectively. Observed discrepancies suggest that the strong-coupling and retardation effects play important role in the analyzed superconducting phase. In this context, presented here results supplements our previous investigations on BKBO superconductor \cite{szczesniak1} and account for the strong-coupling phonon-mediated character of the superconducting phase in BKBO at any doping level.

\begin{acknowledgments}
The Authors would like to note that partial numerical calculations have been conducted on the Cz{\c e}stochowa University of Technology cluster, built in the framework of the PLATON project no. POIG.02.03.00-00-028/08 - the service of the campus calculations U3.
\end{acknowledgments}

\bibliographystyle{apsrev}
\bibliography{bibliography}
%
\end{document}